# Water Filtration Using Plant Xylem

Jongho Lee[†], Michael S. H. Boutilier[†], Valerie Chambers, Varsha Venkatesh, and Rohit Karnik[*]

Department of Mechanical Engineering, Massachusetts Institute of Technology, 77 Massachusetts Avenue, Cambridge MA 02139, USA.

[†]*These authors contributed equally.* [*]*Correspondence: karnik@mit.edu*

**Abstract**

Effective point-of-use devices for providing safe drinking water are urgently needed to reduce the global burden of waterborne disease. Here we show that plant xylem from the sapwood of coniferous trees – a readily available, inexpensive, biodegradable, and disposable material – can remove bacteria from water by simple pressure-driven filtration. Approximately 3 cm$^3$ of sapwood can filter water at the rate of several liters per day, sufficient to meet the clean drinking water needs of one person. The results demonstrate the potential of plant xylem to address the need for pathogen-free drinking water in developing countries and resource-limited settings.

**Introduction**

The scarcity of clean and safe drinking water is one of the major causes of human mortality in the developing world. Potable or drinking water is defined as having acceptable quality in terms of its physical, chemical, and bacteriological parameters so that it can be safely used for drinking and cooking.[1] Among the water pollutants, the most deadly ones are of biological origin: infectious diseases caused by pathogenic bacteria, viruses, protozoa, or parasites are the most common and widespread health risk associated with drinking water.[1] The most common water-borne pathogens are bacteria (e.g. *Escherichia coli*, *Salmonella typhi*, *Vibrio cholerae*), viruses (e.g. adenoviruses, enteroviruses, hepatitis, rotavirus), protozoa (e.g. giardia).[1] These pathogens cause child mortality and also contribute to malnutrition and stunted growth of children. The World Health Organization reports[2] that 1.6 million people die every year from diarrheal diseases attributable to lack of access to safe drinking water and basic sanitation. 90% of these are children under the age of 5, mostly in developing countries. Multiple barriers including prevention of contamination, sanitation, and disinfection are necessary to effectively prevent the spread of waterborne diseases.[1] However, if only one barrier is possible, it has to be disinfection unless evidence exists that chemical contaminants are more harmful than the risk from ingestion of microbial pathogens.[1]

Common technologies for water disinfection include chlorination, filtration, UV-disinfection, pasteurization or boiling, and ozone treatment.[1] Chlorine treatment is effective on a large scale, but becomes expensive for smaller towns and villages. Controlling water quality at the point-of-use is often more effective due to the issues of microbial regrowth, byproducts of disinfectants, pipeline corrosion, and contamination in the distribution system.[3] Boiling is an effective method to disinfect water; however, the amount of fuel required to disinfect water by boiling is several times more than what a typical family will use for cooking.[1] UV-disinfection is the most promising point-of-use technology available,[1] yet it does require access to electricity and some maintenance of the UV lamp, or sufficient sunlight. While small and inexpensive filtration devices can potentially address the issue of point-of-use disinfection, an ideal technology does not currently exist: Inexpensive household carbon-based filters are not capable of removing pathogens and can be used only when the water is already biologically safe.[1] Sand filters that can remove pathogens require large area and knowledge of how to maintain them[1], while membrane filters capable of removing pathogens[3] suffer from high costs, fouling, and require pumping power due to low flow rates[4] that prevents their wide implementation in developing countries. In this context, new approaches that can improve upon current technologies are urgently needed. Specifically, membrane materials that are inexpensive, readily available, disposable, and effective at pathogen removal could greatly impact our ability to provide safe drinking water to the global population.

If we look to nature for inspiration, we find that a potential solution exists in the form of plant xylem – a porous material that conducts fluid in plants.[5] Plants have evolved specialized xylem tissues to conduct sap from their roots to their shoots. Xylem has evolved under the competing pressures of offering minimal resistance to the ascent of sap while maintaining small nanoscale pores to prevent cavitation. The size distribution of these pores – typically a few nanometers to a maximum of around 500 nm, depending on the plant species[6] – also happens to be ideal for filtering out pathogens, which raises the interesting question of whether plant xylem can be used to make inexpensive water filtration devices. Although scientists have extensively studied plant xylem and the ascent of sap, use of plant xylem in the context of water filtration remains to be explored. Measurements of sap flow in plants suggest that flow rates in the range of several liters per hour may be feasible with less than 10 cm-sized filters, using only gravitational pressure to drive the flow.[5]

We therefore investigated whether plant xylem could be used to create water filtration devices. We first reason which type of plant xylem tissue is most suitable for filtration. We then construct a simple water filter from plant xylem and study the resulting flow rates and filtration characteristics. Finally, we show that the xylem filter can effectively remove bacteria from water and discuss directions for further development of these filters.

**Materials and Methods**

*Materials*

Branches were excised from white pine growing in Massachusetts, USA, and placed in water. The pine was identified as *pinus strobus* based on the 5-fold grouping of its needles, the average needle length of 4.5 inches, and the cone shape. Deionized water (Millipore) was used throughout the experiments unless specified otherwise. Red pigment (pigment-based carmine drawing ink, Higgins Inks) was dissolved in deionized water. Nile-red coated 20 nm fluorescent polystyrene nanospheres were obtained from Life Technologies. Inactivated, Alexa 488 fluorescent dye labeled *Escherichia coli* were obtained from Life Technologies. Wood sections were inserted into the end of 3/8 inch internal diameter PCV tubing, sealed with 5 Minute Epoxy, secured with hose clamps, and allowed to cure for ten minutes before conducting flow rate experiments.

*Construction of the Xylem Filter*

1 inch-long sections were cut from a branch with approximately 1 cm diameter. The bark and cambium were peeled off, and the piece was mounted at the end of a tube and sealed with epoxy.

The filters were flushed with 10 mL of deionized water before experiments. Care was taken to avoid drying of the filter.

*Filtration and Flow Rate Experiments*

Approximately 5 mL of deionized water or solution was placed in the tube. Pressure was supplied using a nitrogen tank with a pressure regulator. For filtration experiments, 5 psi pressure was used. The filtrate was collected in glass vials. For dye filtration, size distribution of the pigments was measured for the input solution and the filtrate. Higgins pigment-based carmine drawing ink, diluted ~1000x in deionized water, was used as the input dye solution. For bacteria filtration, the feed solution was prepared by mixing 0.08 mg of inactivated *Escherichia coli* in 20 mL of deionized water (~1.6 x $10^7$ mL$^{-1}$) after which the solution was sonicated for 1 min. The concentration of bacteria was measured in the feed solution and filtrate by enumeration with a hemacytometer (inCyto C-chip) mounted on a Nikon TE2000-U inverted epifluorescence microscope. Before measurement of concentration and filtration experiments, the feed solution was sonicated for 1 min and vigorously mixed.

*Imaging*

Xylem structure was visualized in a scanning electron microscope (SEM, Zeiss Supra55VP). Samples were coated with gold of 5 nm thickness before imaging. To visualize bacteria filtration, 5 mL of solution at a bacteria concentration of ~1.6 x $10^7$ mL$^{-1}$ was flowed into the filter. The filter was then cut longitudinally with a sharp blade. One side of the sample was imaged using a Nikon TE2000-U inverted epifluorescence microscope and the other was coated with gold and imaged with the SEM. Optical images were acquired of the cross section of a filter following filtration of 5 mL of the dye at a dilution of ~ 100x.

*Particle Sizing*

Dynamic light scattering measurements of particle size distributions were performed using a Malvern Zetasizer Nano-ZS.

**Results**

*Xylem Structure and Rationale for Use of Conifer Xylem*

The flow of sap in plants is driven primarily by transpiration from the leaves to the atmosphere, which creates negative pressure in the xylem. Therefore, xylem evolution has occurred under competing pressures of providing minimal resistance to the flow of sap, while protecting against cavitation (i.e. nucleation) and growth of bubbles that could stop the flow of sap and kill the plant, and to do this while maintaining mechanical strength.[5] The xylem structure comprises many small conduits that work in parallel and operate in a manner that is robust to cavitation[5, 6] (Figure 1). In woody plants, the xylem tissue is called the sapwood, which often surrounds the heartwood (i.e. inactive, non-conducting lignified tissue found in older branches and trunks) and is in turn surrounded by the bark (Figure 1b,c). The xylem conduits in gymnosperms (conifers) are formed from single dead cells and are called tracheids (Figure 1c), with the largest tracheids reaching diameters up to 80 μm and lengths up to 10 mm.[5] Angiosperms (flowering plants) have xylem conduits called vessels that are derived from several cells in a single file, having diameters up to 0.5 mm and lengths ranging from a few millimeters to several meters.[5] These parallel conduits have closed ends and are connected to adjacent conduits via "pits"[6] (Figure 1d,e). The pits have membranes with nanoscale pores that perform the critical function of preventing bubbles from crossing over from one conduit to another. Pits occur in a variety of configurations; Figure 1d,e show torus-margo pit membranes that are shaped like a donut (margo) with an impermeable part in the center called torus, occurring in conifers.[6] More interestingly, the porosity of the pit membranes ranges in size from a few nanometers to a few hundred nanometers, with pore sizes in the case of angiosperms tending to be smaller than those in gymnosperms.[6,7] Pit membrane pore sizes have been estimated by examining whether gold colloids or particles of different sizes can flow through.[6,8] Remarkably, it was observed that 20 nm gold colloids could not pass through inter-vessel pit membranes of some deciduous tree species,[8] indicating an adequate size rejection to remove viruses from water. Furthermore, inter-tracheid pit membranes were found to exclude particles in the 200 nm range,[6] as required for removal of bacteria and protozoa.

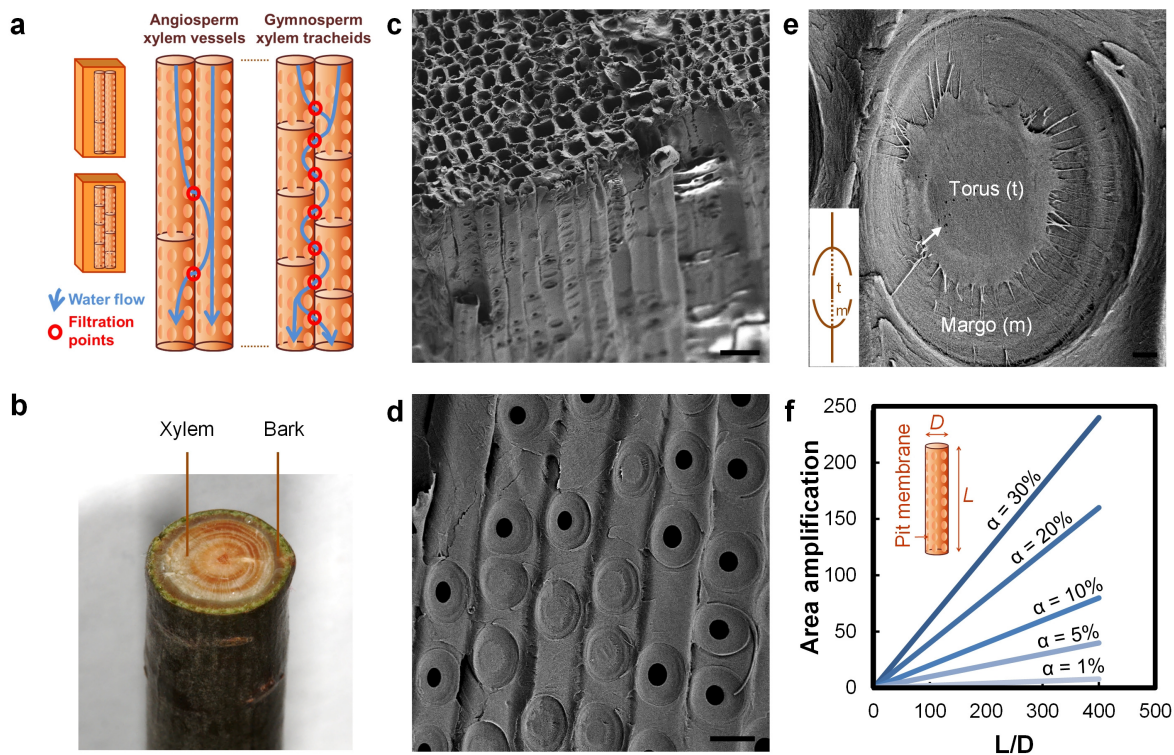

**Figure 1.** Xylem structure. a) Structure of xylem vessels in flowering plants and tracheids in conifers. Longer length of the vessels can provide pathways that can bypass filtration through pit membranes that decorate their circumference. b) Photograph of ~ 1 cm diameter pine (*pinus strobus*) branch used in the present study. c) Scanning electron microscope (SEM) image of cut section showing tracheid cross section and lengthwise profile. Scale bar is 40 μm. d) SEM image showing pits and pit membranes. Scale bar is 20 μm. e) Pit membrane with inset showing a cartoon of the pit cross-section. The pit cover has been sliced away to reveal the permeable margo surrounding the impermeable torus. Arrow indicates observed hole-like structures that may be defects. The margo comprises radial spoke-like structures that suspend the torus that are only barely visible overlaying the cell wall in the background. Scale bar 1 μm. f) Dependence of the area amplification, defined as the pit membrane area divided by the nominal filter area, on the tracheid aspect ratio $L/D$ and fractional area $\alpha$ occupied by pit membranes.

Since angiosperms (flowering plants, including hardwood trees) have larger xylem vessels that are more effective at conducting sap, xylem tissue constitutes a smaller fraction of the cross-section area of their trunks or branches, which is not ideal in the context of filtration. The long length of their xylem vessels also implies that a large thickness (centimeters to meters) of xylem tissue will be required to achieve any filtration effect at all – filters that are thinner than the average vessel length will just allow water to flow through the vessels without filtering it through pit membranes (Figure 1a). In contrast, gymnosperms (conifers, including softwood trees) have short tracheids that would force water to flow through pit membranes even for small thicknesses (< 1 cm) of xylem tissue (Figure 1a). Since tracheids have smaller diameters and are shorter, they offer higher resistance to flow, but typically a greater fraction of the stem cross-section area is devoted to conducting xylem tissue. For example, in the pine branch shown in Figure 1b used in this study, fluid-conducting xylem constitutes the majority of the cross-section. This reasoning leads us to the conclusion that in general the xylem tissue of coniferous tress – i.e. the sapwood – is likely to be the most suitable xylem tissue for construction of a water filtration device, at least for filtration of bacteria, protozoa, and other pathogens on the micron or larger scale.

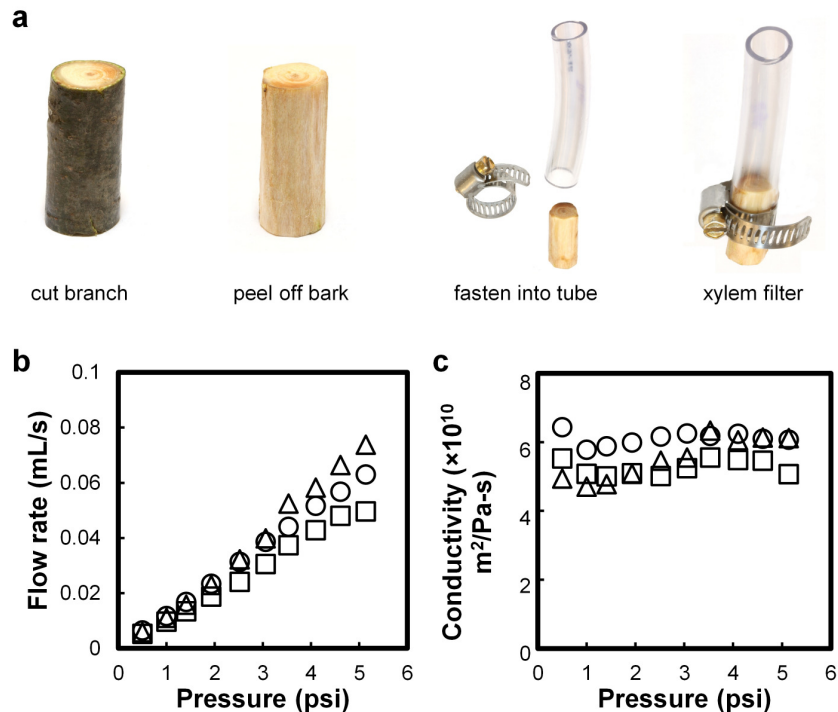

**Figure 2.** Xylem filter. a) Construction of xylem filter. b) Effect of applied pressure on the water flow rate through the xylem filter. c) Hydrodynamic conductivity of the filter extracted at each measured pressure using the total filter cross-section area and length as defined by Equation 1.

The resistance to fluid flow is an important consideration for filtration. Pits can contribute a significant fraction (as much as 30-80%)[5, 6] of the resistance to sap flow, but this is remarkably small considering that pit membrane pore sizes are several orders of magnitude smaller than the tracheid or vessel diameter. The pits and pit membranes form a hierarchical structure where the thin, highly-permeable pit membranes are supported across the microscale pits that are arranged around the circumference of the tracheids (Figure 1a). This arrangement permits the pit membranes to be thin, offering low resistance to fluid flow. Furthermore, the parallel arrangement of tracheids with pits around their circumference provides a high packing density for the pit membranes. For a given tracheid with diameter $D$ and length $L$, where pit membranes occupy a fraction $\alpha$ of the tracheid wall area, each tracheid effectively contributes a pit membrane area of $\pi D L \alpha / 2$, where the factor of 2 arises as each membrane is shared by two tracheids. However, the nominal area of the tracheid is only $\pi D^2/4$, and therefore, the structure effectively amplifies the nominal filter area by a factor of $2\alpha(L/D)$ (Figure 1f). The images in Figure 1c indicate $D \sim 10\text{-}15$ μm, $\alpha \sim 0.2$, yielding an effective area amplification of ~20 for tracheid lengths of 1-2 mm. Therefore, for a filter made by cutting a slice of thickness $\sim L$ of the xylem, the actual membrane area is greater by a large factor due to vertical packing of the pit membranes. Larger filter thicknesses further increase the total membrane area, but the additional area of the membrane is positioned in series rather than in parallel and therefore decreases the flow rate, but potentially improves the rejection performance of the filter.

*Construction of the Xylem Filter and Measurement of Flow Rate*

The xylem filter device was constructed by simply peeling off the bark from a section of the pine branch and inserting it into a tube (Figure 2a). Although a simple tube fastener could provide a leak-tight seal between the tube and the xylem, we used epoxy to ensure that there was no inadvertent leakage. When deionized water was loaded into the tube above the xylem and subjected to pressure in the 0.5-5 psi range, we found that water readily flowed through the xylem. The flow rate was proportional to applied pressure (Figure 2b), which allowed for the extraction of the hydrodynamic conductivity $K$ (m$^2$/Pa-s) of the filter, defined by

$$Q = KA\frac{\Delta P}{l} \tag{1}$$

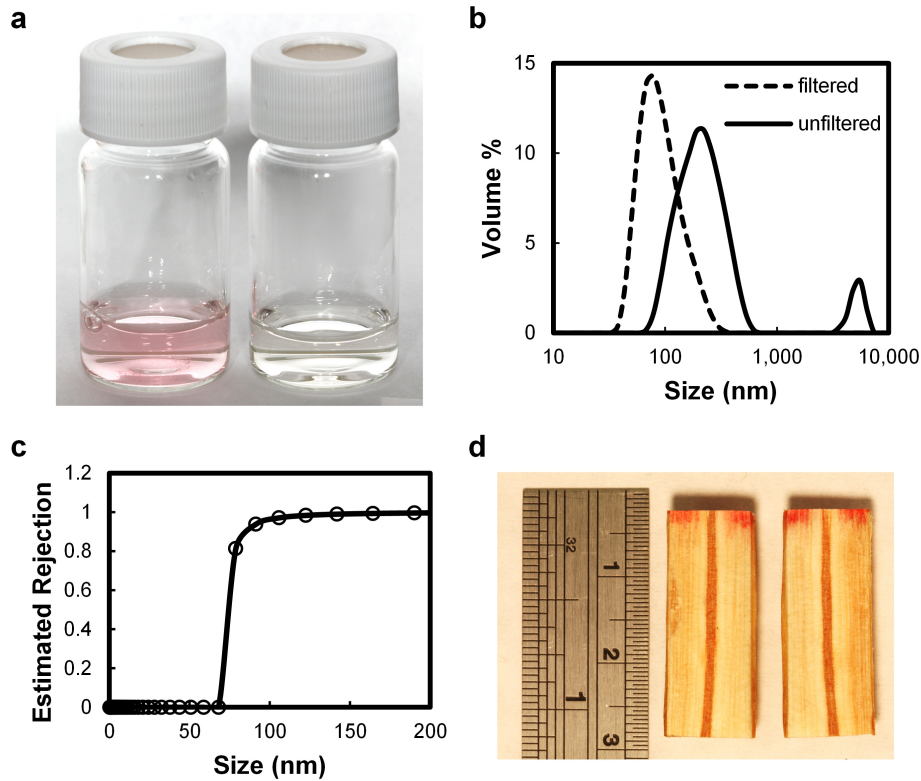

**Figure 3.** Filtration performance of the xylem filter. a) Feed solution of a pigment dye before filtration, compared to the filtrate. b) Size distribution of the pigment particles in the feed and filtrate solutions measured by dynamic light scattering. c) Dependence of the rejection on the particle size estimated from the data in (c). d) Cross-section of the xylem filter after filtration. Scale is in centimeters and inches.

where $Q$ is the volumetric flow rate (in m$^3$/s) under pressure difference $\Delta P$ across the filter, while $l$ and $A$ are the thickness and the cross-section area of the filter, respectively. The observed conductivities for three different filters were in the range of ~5-6×10$^{-10}$ m$^2$/ Pa-s (Figure 2c), or equivalently, ~0.5-0.6 kg s$^{-1}$ m$^{-1}$ MPa$^{-1}$ when defined with respect to mass flow rate of water.

Biologists have performed similar flow rate measurements by cutting a section of a plant stem under water, flushing to remove any bubbles, and applying a pressure difference to measure the flow rate.[9, 10] Xylem conductivities of conifers[5] typically range from 1-4 kg s$^{-1}$ m$^{-1}$ MPa$^{-1}$, which compares very well with the conductivities measured in our experiments. Lower conductivities can easily result from introduction of bubbles[9] or the presence of some non-conducting heartwood. We can therefore conclude that the flow rate measurements in our devices are consistent with those expected from prior literature on conductivity of conifer xylem.

*Filtration of Pigment Dye*

After construction of the filter, we tested its ability to filter a pigment dye with a broad particle size distribution. The red color of the feed solution disappeared upon filtration (Figure 3a) indicating that the xylem filter could effectively filter out the dye.

Since the dye had a broad pigment size distribution, we investigated the size-dependence of filtration by quantifying the pigment size distribution before and after filtration using dynamic light scattering. We found that the feed solution comprised particles ranging in size from ~70 nm to ~500 nm, with some larger aggregates (Figure 3b). In contrast, the filtrate particle size distribution peaked at that ~80 nm, indicating that larger particles were filtered out. In a separate experiment, we observed that 20 nm fluorescent polystyrene nanoparticles could not be filtered by the xylem filter, confirming this size dependence of filtration. Assuming that pigment particles 70 nm or less in size were not rejected, the size distributions before and after filtration enable calculation of the rejection performance of the xylem filter as a function of particle size (Figure 3c). We find that the xylem filter exhibits excellent rejection for particles with diameters exceeding 100 nm, with the estimated rejection exceeding 99% for particles over 150 nm. Smaller particles are expected to pass through the larger porosity of the pit membrane: SEM images in Figure 1e indicate sub-micron spacing between the radial spoke-like margo membrane through which the pigment particles can pass, although the porosity is difficult to resolve.

After filtration, we cut the xylem filter parallel to the direction of flow to investigate the distribution of dye in the filter. We observed that the dye was confined to the top 2-3 millimeters of the xylem filter (Figure 3d), which compares well with the tracheid lengths on the millimeter scale expected for coniferous trees.[5] These results show that the majority of the filtration occurred within this length scale, and suggests that the thickness of the xylem filter may be decreased to below 1 cm while still rejecting the majority of the dye.

*Filtration of Bacteria from Water*

Finally, we investigated the ability of the xylem filter to remove bacteria from water. As a model bacterium, we used fluorescently labeled and inactivated *Escherichia coli* bacteria that are cylindrical in shape with a diameter of ~1 μm. Use of fluorescently labeled particles enabled easy enumeration of their concentrations, as well as allowed for tracking of the location in the xylem filter where they were trapped. Filtration using three different xylem filters showed nearly complete rejection of the bacteria (Figure 4a). Using a hemacytometer to count the bacteria, we estimate that the rejection was at least 99.9%.

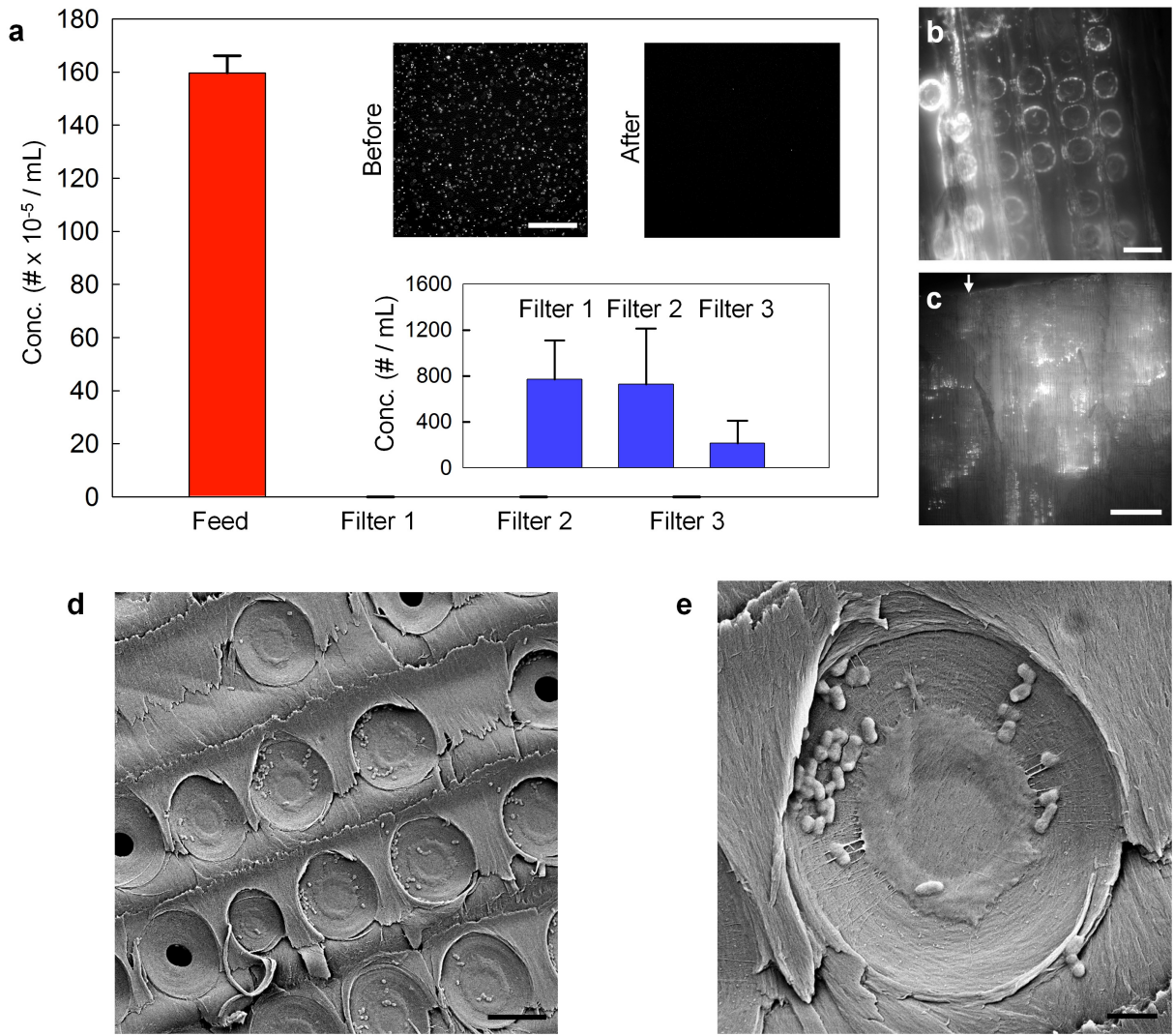

**Figure 4.** Filtration of model bacteria by the xylem filter. a) Concentrations of bacteria in the feed and filtrate solutions. Inset shows fluorescence images of the two solutions. Scale bar is 200 μm. b) Fluorescence image of xylem filter cross-section showing accumulation of bacteria over the margo pit membranes. Scale bar is 20 μm. c) Low-magnification fluorescence image shows that bacteria are trapped at the bottoms of tracheids within the first few millimeters of the top surface. Scale bar is 400 μm. Arrow indicates top surface of the xylem filter and also the direction of flow during filtration. Autofluorescence of the xylem tissue also contributes to the fluorescence signal in (b) and (c). d,e) SEM images showing bacteria accumulated on the margo pit membranes after filtration. Scale bars are 10 μm and 2 μm, respectively.

To investigate the mechanism of filtration, after filtration we cut the xylem filter parallel to the direction of flow. When examined under a fluorescence microscope, we observed the bacteria accumulated over the donut-shaped margo pit membranes (Figure 4b). This distribution is consistent with the expectation that the bacteria are filtered by the pit membranes. The distribution of trapped bacteria was not uniform across the cross section. Similar to the case of the dye, bacteria were observed only within the first few millimeters from the end through which the solution was infused (indicated by white arrow in Figure 4c). In addition, the low-magnification fluorescence image suggests that the bacteria may accumulate primarily over pit membranes at the bottom of the tracheids, which is again not unexpected. Further investigation by SEM confirmed that the bacteria had accumulated on the pit membranes (Figure 4d,e). These results confirm the pit membranes as the functional units that provide the filtration effect in the xylem filter.

**Discussion**

Wood has been investigated recently as a potential filtration material,[11] showing moderate improvement of turbidity. While we showed filtration using freshly cut xylem, we found that the flow rate dropped irreversibly by over a factor of 100 if the xylem was dried, even when the xylem was flushed with water before drying. We also examined flow through commercially available kiln-dried wood samples cut to similar dimensions. We found that wood samples that exhibited filtration also had flow rates that were two orders of magnitude smaller than in the fresh xylem filter, and those that had high flow rates did not exhibit much filtration effect and seemed to have ruptured tracheids and membranes when observed under SEM. Wetting using ethanol or vacuum to remove air did not significantly increase flow rate, suggesting that the pit membranes may have a tendency to become clogged during drying. These results are consistent with literature showing that the pit membranes can become irreversibly aspirated against the cell wall, blocking the flow.[12] In fact, the pit membranes in the SEM images (Figure 1d,e and Figure 4d,e), which were acquired after drying the samples, appear to be stuck to the walls. Regardless, our results demonstrate that excellent rejection (>99.9%) of bacteria is possible using the pit membranes of fresh plant xylem, and also provide insight into the mechanism of filtration as well as guidelines for selection of the xylem material.

The vertical arrangement of the tracheids makes them susceptible to concentration polarization and fouling (i.e. accumulation of the rejected particles in the tracheids), and therefore xylem filters are likely to be most advantageous for removal of pathogens from water that is not significantly turbid. The pressures of 1-5 psi used here are easily achievable using a gravitational pressure head of 0.7 – 3.5 m, implying that no pumps are necessary for filtration. The measured flow rates of about 0.05 mL/s using only ~ 1 $cm^2$ filter area correspond to a flow rate of over 4 L/d, sufficient to meet the drinking water requirements of one person. While membranes are easily fouled, xylem filters could be easily replaced due to their biodegradability and low cost.

Wood is an easily available material. While use of fresh xylem does not preclude its use as a filter material, dried membranes have definite practical advantages. Therefore, the process of wood drying and its influence on xylem conductivity needs further study. In particular, processes that yield intact yet permeable xylem tissues that can be stored long-term will be helpful for improving the supply chain if these filters are to be widely adopted. In addition, flow through xylem of different plants needs to be studied to identify locally available sources of xylem, which will truly enable construction of filters from locally available materials. In the present study, we report results using xylem derived from only one species. These xylem filters could not filter out 20 nm nanoparticles, which is a size comparable to that of the smallest viruses. It will be interesting to explore whether there exist any coniferous species that have pit membranes with smaller pore sizes that can filter out viruses. In their absence, angiosperms with short vessels that yield practical filter lengths may be the best alternative due to their smaller pit membrane pore sizes.[6] Further exploration of xylem tissues from different plants with an engineering perspective is needed to construct xylem filters that can effectively reject viruses, exhibit improved flow rates, or that are amenable to easy storage.

**Conclusions**

Plant xylem is a porous material with membranes comprising nanoscale pores. We have reasoned that xylem from the sapwood of coniferous trees is suitable for disinfection by filtration of water. The hierarchical arrangement of the membranes in the xylem tissue effectively amplifies the available membrane area for filtration, providing high flow rates. Xylem filters were prepared by simply removing the bark of pine tree branches and inserting the xylem tissue into a tube. Pigment filtration experiments revealed a size cutoff of about 100 nm, with most of the filtration occurring within the first 2-3 mm of the xylem filter. The xylem filter could effectively filter out bacteria from water with rejection exceeding 99.9%. Pit membranes were identified as the functional unit where actual filtration of the bacteria occurred. Flow rates of about 4 L/d were obtained through ~ 1 cm$^2$ filter areas at applied pressures of about 5 psi, which is sufficient to meet the drinking water needs of one person. The simple construction of xylem filters, combined with their fabrication from an inexpensive, biodegradable, and disposable material suggests that further research and development of xylem filters could lead to their widespread use and greatly reduce the incidence of waterborne infectious disease in the world.

**Acknowledgements**

This work was supported by the James H. Ferry, Jr. Fund for Innovation in Research Education award to R.K. for the proposal entitled "Can Plant Xylem Provide Safe Drinking Water?" SEM imaging was performed at the Harvard Center for Nanoscale Systems, a member of the National Nanotechnology Infrastructure Network (NNIN), which is supported by the National Science Foundation under NSF award no. ECS-0335765. The authors thank Yukiko Oka for assistance with preparation of illustrations and Sunandini Chopra for help with dynamic light scattering measurements.